%% file: eprint_dpf2013.tex
\documentclass[12pt]{article}
\usepackage{graphicx}

\newcommand\pubnumber{DPF2013-308}
\newcommand\pubdate{\today}

\def\colorado{Department of Physics\\
University of Colorado, Boulder, CO 80309-0390}
\def\support{\footnote{keith.ulmer@colorado.edu}}

\def\Title#1{\begin{center} {\Large #1 } \end{center}}
\def\Author#1{\begin{center}{ \sc #1} \end{center}}
\def\Address#1{\begin{center}{ \it #1} \end{center}}

\newcommand\pubblock{\rightline{\begin{tabular}{l} \pubnumber\\
         \pubdate  \end{tabular}}}
\newenvironment{Abstract}{\begin{quotation}  }{\end{quotation}}
\newenvironment{Presented}{\begin{quotation} \begin{center} 
             PRESENTED AT\end{center}\bigskip 
      \begin{center}\begin{large}}{\end{large}\end{center} \end{quotation}}

\input econfmacros.tex

\begin{document}
\begin{titlepage}
\pubblock

\vfill
\Title{Future Sensitivity Studies for Supersymmetry Searches at CMS at 14 TeV}
\vfill
\Author{ Keith A. Ulmer\support}
\Address{\colorado}
\Author{on behalf of the CMS Collaboration}
\vfill
\begin{Abstract}
The sensitivity for CMS searches for supersymmetry is evaluated in the context of an upgraded LHC 
at a center-of-mass energy of 14 TeV and an integrated luminosity of 300 $\textrm{fb}^{-1}$. 
Results for several key searches for supersymmetry are presented including direct and 
gluino-mediated stop and sbottom production and electroweak production of supersymmetric particles.
\end{Abstract}
\vfill
\begin{Presented}
DPF 2013\\
The Meeting of the American Physical Society\\
Division of Particles and Fields\\
Santa Cruz, California, August 13--17, 2013\\
\end{Presented}
\vfill
\end{titlepage}
\def\thefootnote{\fnsymbol{footnote}}
\setcounter{footnote}{0}

\section{Introduction}

In the first three years of operation from 2010 to 2012 the Large Hadron Collider (LHC)
at CERN and the Compact Muon Solinoid (CMS)~\cite{CMS} experiment have performed remarkably
well. A total of $\sim 25 \textrm{fb}^{-1}$ of data have been collected at center-of-mass
energies up to 8 TeV, the highest ever achieved in a particle collider. After the discovery
of the Higgs Boson~\cite{CMShiggs,ATLAShiggs}, one of the most pressing questions to
address at the LHC is what mechanism suppresses the quantum divergences that would
appear as corrections to the Higgs mass. A natural solution to this hierarchy problem
has justifiably become a key focus of current and future studies at the LHC.

A wide range of searches for beyond the standard model (BSM) physics has been performed
at CMS and ATLAS, thus far with no evidence for BSM physics. The window for a natural solution
to the hierarchy problem is near, with the data taken at the LHC after the current long shutdown
expected to play a key role in discovering or ruling out such a scenario.

Projections for future sensitivity with up to $300 \textrm{fb}^{-1}$ at $\sqrt{s} = 14$
TeV are presented here from several key searches for supersymmetry (SUSY) from the CMS experiment. 
Existing searches at 8 TeV are extrapolated for the projections by scaling the
luminosity and taking into account the change in cross section with higher energy for
signal and background. The projections are made based on 8 TeV Monte Carlo simulation
samples and without optimizing the selections for searches at higher energies and 
high luminosities. The results may therefore be taken as conservative estimates as the
true searches done at 14 TeV will be reoptimized.

The studies presented here were prepared as part of the CMS contribution to the
Snowmass process~\cite{snowmass} to assess the long-term physics
aspirations of the US high energy physics community. More details about these
results and other CMS contributions to Snowmass may be found in the submitted
whitepaper~\cite{whitepaper}.

\section{Squark discovery potential}

A natural solution to the hierarchy problem is possible from supersymmetry if the
partner of the top quark, the stop squark, provides a cancellation to the large
radiative correction to the Higgs mass induced by top quark loops. To avoid fine
tuning, the mass of the stop squark should be less than $\sim 1$ TeV. Performance
in stop squark searches is benchmarked with the simplified model (SMS)~\cite{SMS} of
pair production of stops where each decays to a top quark and the lightest supersymmetric
particle (LSP) with 100$\%$ branching fraction. All other sparticles are
decoupled.

\begin{figure}[htb]
\centering
\includegraphics[width=0.48\linewidth]{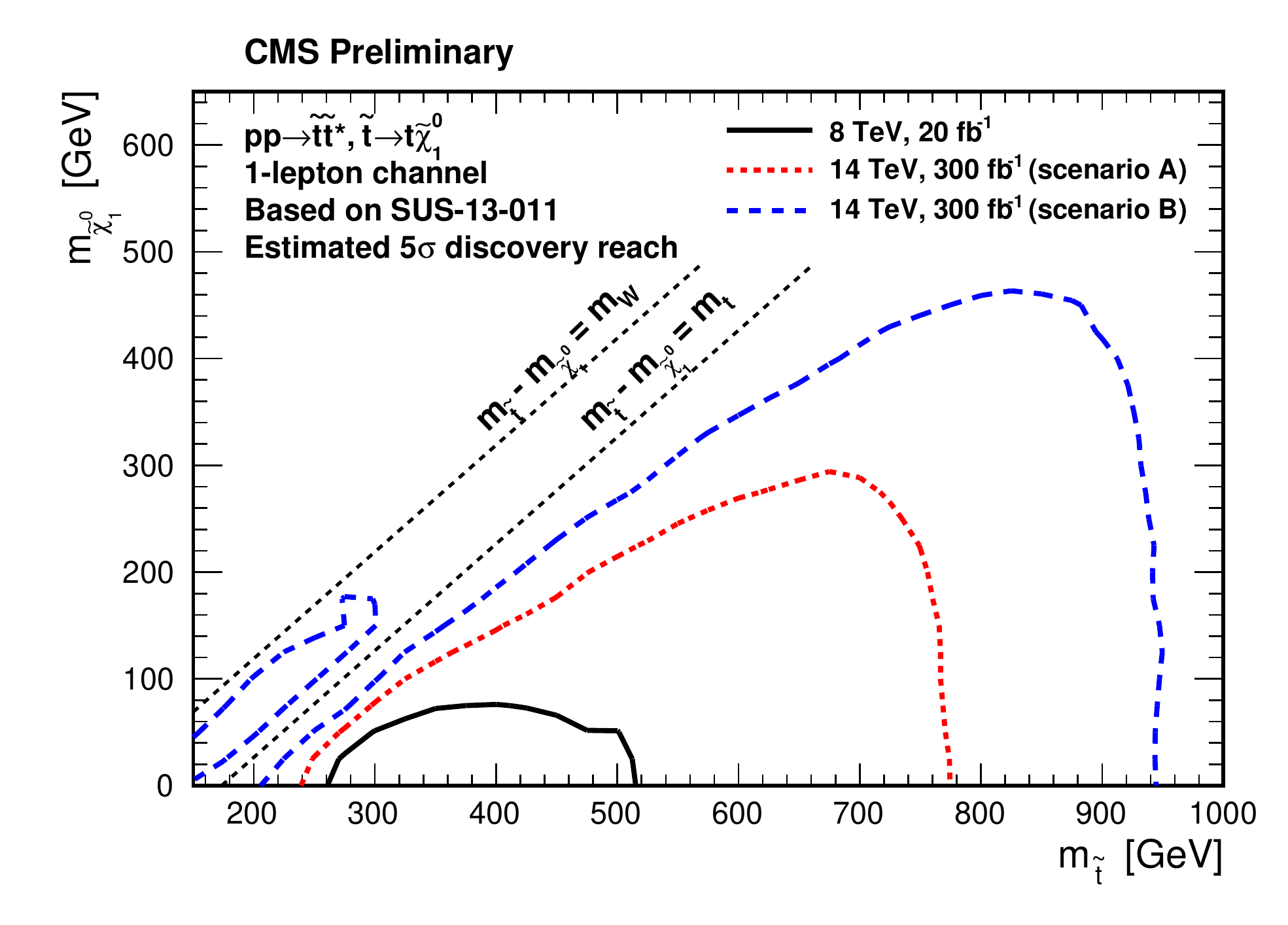}
\includegraphics[width=0.48\linewidth]{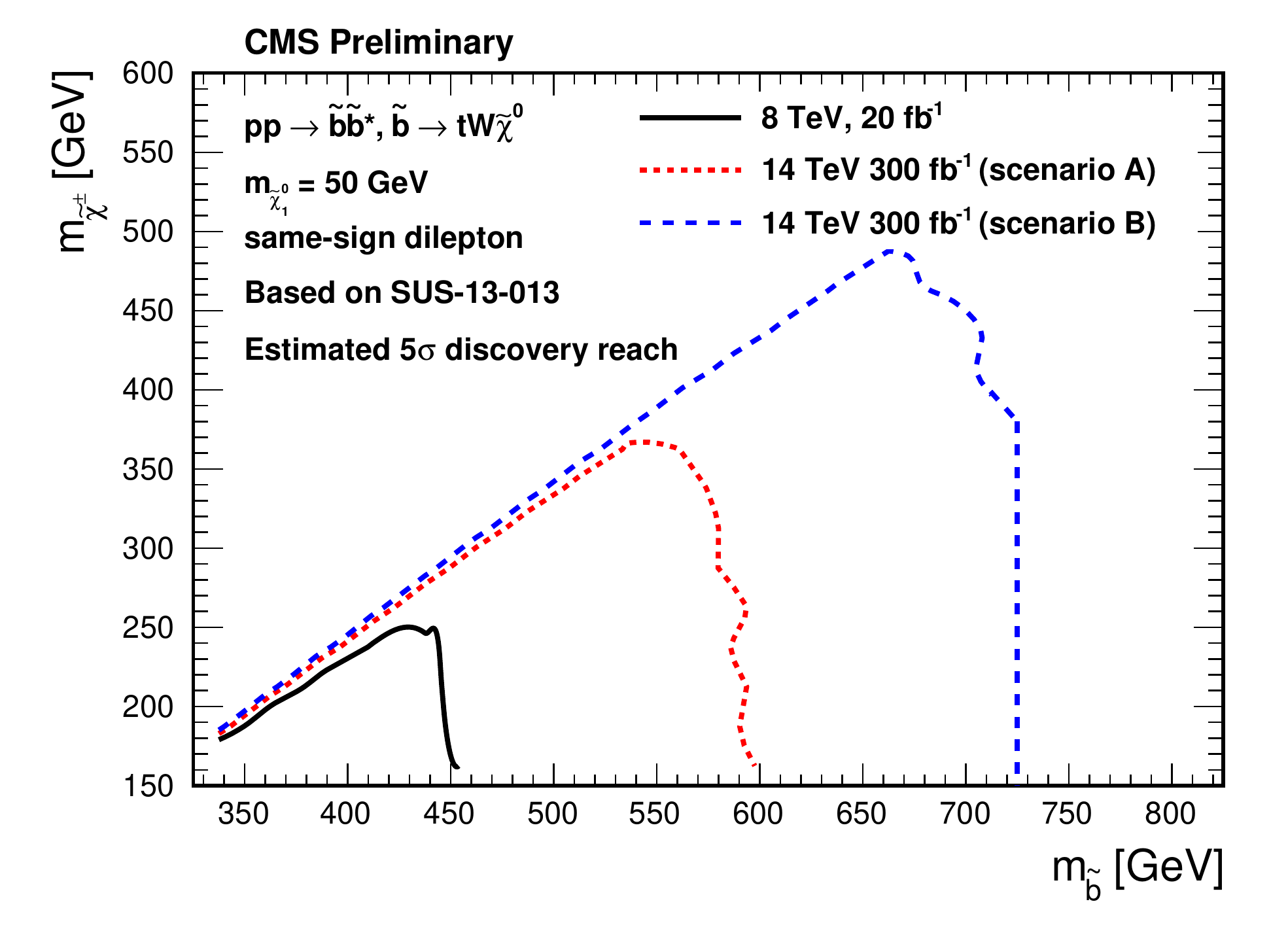}
\caption{Projected 5$\sigma$ discovery reach for stop to top, LSP (left)
and sbottom to top, chargino (right) simplified models.}
\label{fig:squark}
\end{figure}

Future projections are based on the extrapolation of a CMS search at 8 TeV in
final states with a single muon or electron~\cite{stop}. This analysis has a
discovery reach for stop masses 300-500 GeV and a maximum neutralino (LSP) mass of
75 GeV for center-of-mass energy of 8 TeV and an integrated luminosity of 20 
$\textrm{fb}^{-1}$. The expected event counts for signal and background are predicted
by scaling the
luminosity and taking into account the change in cross section with higher energy.
The systematic uncertainty is considered in two scenarios. In ``scenario A'' the 
systematic uncertainty on the background is scaled from the 8 TeV search by
the ratio of the luminosities and cross sections, as is done for the signal and 
background yields. In the less conservative ``scenario B'' the uncertainty on the
background is reduced by an additional factor of the square root of the ratio of
the luminosity and cross section for 8 and 14 TeV. Scenario B is based on the 
assumption that increased statistics in the data control regions will allow for
better precision of the background estimation with additional data. The results
for both scenarios can be seen in the left plot of Fig.~\ref{fig:squark} where the 
expected mass reach as functions of stop and LSP mass are shown for potential 5$\sigma$
discoveries.

In a natural SUSY scenario, the partner of the bottom quark, the sbottom, is also
expected to be light based on the SU(2) relationship to the stop. Future performance
for sbottom searches is benchmarked in the SMS with pair production of sbottom squarks
where each sbottom decays to a top quark and a chargino, with the chargino subsequently
decaying to a $W$ and the LSP with 100$\%$ branching fraction. The projection is 
extrapolated from the CMS search at 8 TeV for the final state with two same-sign, 
isolated electrons or muons~\cite{sbottom} with signal and background
yields based on scaling up the luminosity and cross section as in the stop search.
Two scenarios are considered for the systematic uncertainties on the background
predictions. The main contributions to the background are from two sources: rare
standard model events with two true same-sign leptons; and events with a non-prompt
lepton, such as from heavy flavor decays, that fakes the isolated lepton signature.
The 8 TeV result is limited by systematic uncertainties on both of these backgrounds
with an uncertainty of 50$\%$ for each. For scenario A the 50$\%$ uncertainty is 
retained. For a less conservative scenario B, the uncertainty on the rare and fake
backgrounds are assumed to be able to be reduced to 30$\%$ and 40$\%$, respectively.
Results for the sbottom projections in both scenarios are shown in the right plot
of Fig.~\ref{fig:squark}.

For both stops and sbottoms, the possible discovery reach is greatly extended for
a center-of-mass energy of 14 TeV with 300 $\textrm{fb}^{-1}$. Stop squark discovery
reaches a potential mass of $\sim 900$ GeV in the optimistic scenario with a massless
LSP, while potential sbottom discoveries are probed up to a sbottom mass of $\sim 725$
GeV. 

\section{Gluino discovery potential}

In addition to light third generation squarks, naturalness also predicts gluinos that
are not much heavier than around 1 TeV. In the scenario with light stops and sbottoms,
and heavier other squarks, gluinos would decay into stops and sbottoms. In this section,
we consider two such gluino decays where the squarks are offshell and represented by
the three body gluino decays into $q$, $q$, LSP, where $q$ represents either a top quark
or a bottom quark.

\begin{figure}[htb]
\centering
\includegraphics[width=0.48\linewidth]{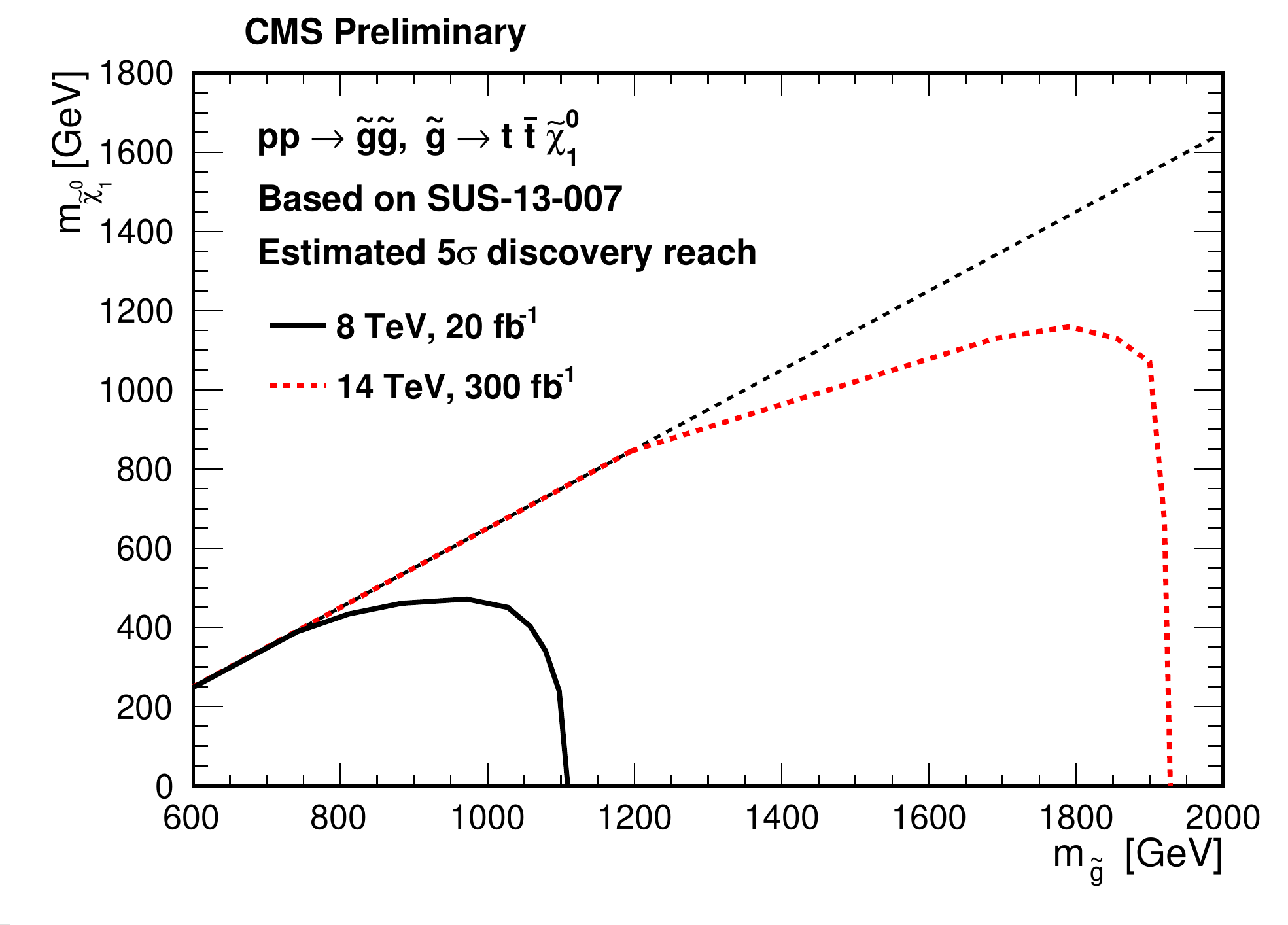}
\includegraphics[width=0.48\linewidth]{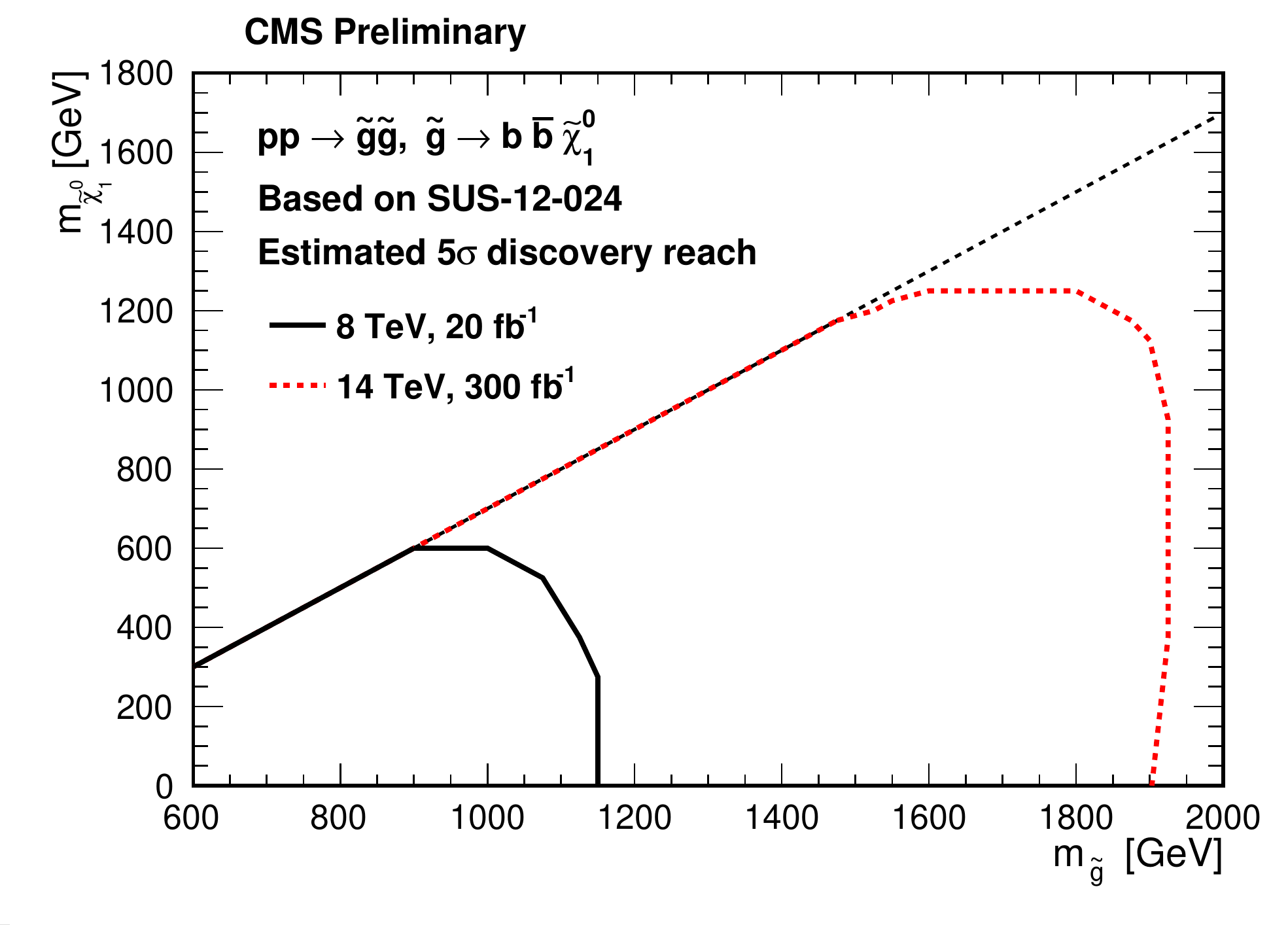}
\caption{Projected 5$\sigma$ discovery reach for gluino to top, top, LSP (left)
and gluino to bottom, bottom, LSP (right) simplified models.}
\label{fig:gluino}
\end{figure}

For the gluino-mediated stop production scenario, an 8 TeV CMS analysis searching for
final states with a single isolated electron or muon~\cite{glstop} is used to project
sensitivity to gluino discovery at 14 TeV and 300 $\textrm{fb}^{-1}$. The numbers of signal
and background events are scaled from the 8 TeV analysis as for the squark results. In this
analysis the dominant uncertainty on the background yield is due to the statistical 
uncertainty of the number of events in relevant data control regions. This uncertainty
will scale with the increase in statistics due do the higher luminosity and higher 
cross section, and therefore is expected to go down as the square root of the ratio
of luminosity and cross section for the two scenarios. Other systematic contributions to
the background yield are of minimal importance, and thus only one scenario is presented
for this study. The 5$\sigma$ discovery sensitivity reach for the projections are shown 
in the left plot of Fig.~\ref{fig:gluino}.

The CMS sensitivity for gluino-mediated sbottom production is studied by extrapolating from
an all hadronic search performed at 8 TeV~\cite{glsbottom}. As for the other projections, the
expected signal and background yields are estimated by scaling the 8 TeV results by the 
increased luminosity and cross sections. As for the gluino-mediated stop search, the
uncertainty on the background yield is dominated by statistical uncertainties in data
control regions, and thus is expected to scale as the square root of the ratio of the 
increased luminosity and cross sections between 8 and 14 TeV. The expected 5$\sigma$ 
discovery reach is shown in the right plot of Fig.~\ref{fig:gluino}. For both of the 
gluino decays studied, the sensitivity is greatly extended for 300 $\textrm{fb}^{-1}$
at 14 TeV with the discovery reach in each case extending beyond gluinos with mass
of 1.9 TeV.

\section{Electroweak SUSY discovery potential}

If gluinos and squarks are too heavy to be produced abundantly at the LHC, direct
production of electroweak SUSY particles may offer the best discovery potential
for supersymmetry. Electroweak production is benchmarked here by an SMS
with direct production of a
neutralino and a chargino ($\tilde{\chi}^{\pm}_1\tilde{\chi}^{0}_2$), where the
neutralino decays into a $Z$ boson and the LSP, while the chargino decays into
a $W$ boson and the LSP, each with 100$\%$ branching fraction.
The future sensitivity for this channel is estimated by projecting an  
8 TeV CMS search for a final state with three or more leptons~\cite{ewkino}.

\begin{figure}[htb]
\centering
\includegraphics[width=0.48\linewidth]{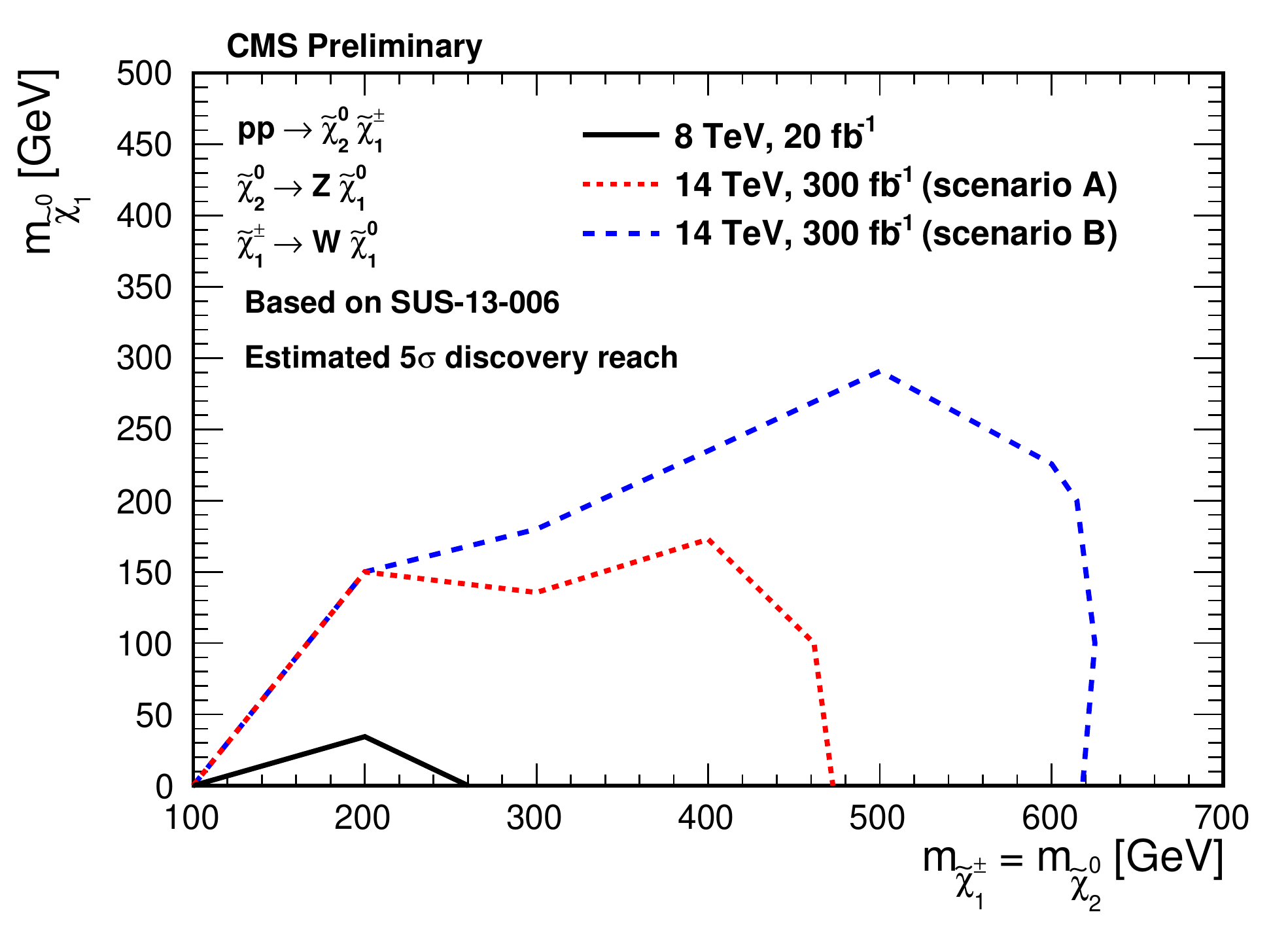}
\caption{Projected 5$\sigma$ discovery reach for direct $\tilde{\chi}^{\pm}_1\tilde{\chi}^{0}_2$ production.}
\label{fig:ewk}
\end{figure}

Expected signal and background yields are projected as for the other searches,
and two scenarios are considered for the uncertainty of the background yield
as described for the direct stop search. The results of the projections for 
expected 5$\sigma$ discovery sensitivity are shown in Fig.~\ref{fig:ewk}. The 
low cross section for electroweak SUSY production does not allow for a significant
discovery reach with the 8 TeV search, while the projection to 300 $\textrm{fb}^{-1}$
at 14 TeV shows that discovery over a significant range will be possible.

\section{Conclusions}

The sensitivity of CMS searches for supersymmetry with an upgraded LHC at a center-of-mass
energy of 14 TeV and 300 $\textrm{fb}^{-1}$ are presented. For each of gluino, squark and 
gaugino production the discovery potential is significantly extended in the upgraded scenario
compared to the current (null) results obtained with the 8 TeV data. In particular, the
region of parameter space of interest for discovering or ruling out a natural 
solution to the hierarchy problem will be probed.

\end{document}

%% file: econfmacros.tex
\def\beq{\begin{equation}}
\def\eeq#1{\label{#1}\end{equation}}
\def\eeqn{\end{equation}}

\def\beqa{\begin{eqnarray}}
\def\eeqa#1{\label{#1}\end{eqnarray}}
\def\eeqan{\end{eqnarray}}

\let\bar=\overbar

\def\Dslash{\not{\hbox{\kern-4pt $D$}}}
\def\dslash{\not{\hbox{\kern-2pt $\del$}}}

\def\msb{{\bar{\ssstyle M \kern -1pt S}}}




%% file: eprint_dpf2013.bbl
\begin{thebibliography}{99}


\bibitem{CMS}
  S.~Chatrchyan {\it et al.}  [CMS Collaboration], ``The CMS experiment at the CERN LHC,'' 
  JINST {\bf 3}, S08004 (2008).

\bibitem{CMShiggs}
  S.~Chatrchyan {\it et al.}  [CMS Collaboration],
  ``Observation of a new boson at a mass of 125 GeV with the CMS experiment at the LHC,''
  Phys.\ Lett.\ B {\bf 716}, 30 (2012)
  [arXiv:1207.7235 [hep-ex]].

\bibitem{ATLAShiggs}
  G.~Aad {\it et al.}  [CMS Collaboration],
  ``Observation of a new particle in the search for the Standard Model Higgs boson with the ATLAS detector at the LHC,''
  Phys.\ Lett.\ B {\bf 716}, 1 (2012)
  [arXiv:1207.7214 [hep-ex]].

\bibitem{snowmass}
See http://www.snowmass2013.org/tiki-index.php.

\bibitem{whitepaper} 
  S.~Chatrchyan {\it et al.}  [CMS Collaboration],
  ``Projected Performance of an Upgraded CMS Detector at the LHC and HL-LHC: Contribution to the Snowmass Process,''
  arXiv:1307.7135 [hep-ex].

\bibitem{SMS}
  D.~Alves {\it et al.}, ``Simplified models for LHC new physics searches,'' arXiv:1105.2838 [hep-ph].

\bibitem{stop} 
  S.~Chatrchyan {\it et al.}  [CMS Collaboration],
  ``Search for top-squark pair production in the single-lepton final state in pp collisions at $\sqrt{s} = 8$ TeV,''
  arXiv:1308.1586 [hep-ex].

\bibitem{sbottom}
The CMS Collaboration, CMS Physics Analysis Summary {\bf SUS-13-013},
``Search for new physics in events with same-sign dileptons and jets in pp collisions at $\sqrt{s}$ = 8 TeV,''
https://twiki.cern.ch/twiki/bin/view/CMSPublic/PhysicsResultsSUS13013, (2013).

\bibitem{glstop}
The CMS Collaboration, CMS Physics Analysis Summary {\bf SUS-13-007},
``Search for supersymmetry in pp collisions at a center-of-mass energy of 8 TeV in events with a single lepton, multiple jets and b-tags,''
https://twiki.cern.ch/twiki/bin/view/CMSPublic/PhysicsResultsSUS13007, (2013).

\bibitem{glsbottom} 
  S.~Chatrchyan {\it et al.}  [CMS Collaboration],
  ``Search for gluino mediated bottom- and top-squark production in multijet final states in pp collisions at 8 TeV,''
  Phys.\ Lett.\ B {\bf 725}, 243 (2013)
  [arXiv:1305.2390 [hep-ex]].
  
\bibitem{ewkino}
The CMS Collaboration, CMS Physics Analysis Summary {\bf SUS-13-006},
``Search for electroweak production of charginos, neutralinos, and sleptons using leptonic final states in pp collisions at $\sqrt{s}$ = 8 TeV,''
https://twiki.cern.ch/twiki/bin/view/CMSPublic/PhysicsResultsSUS13006, (2013).

\end{thebibliography}
